\documentclass[aps,pra,groupedaddress,amsmath,amssymb,10pt,nofootinbib,notitlepage]{revtex4-1}

\usepackage{graphicx}
\usepackage[utf8]{inputenc}
\usepackage{mathtools}

\usepackage{amsmath}
\usepackage{amssymb}
\usepackage{amsthm}

\usepackage{fontenc}

\usepackage[usenames,dvipsnames]{color}
\usepackage[capitalize]{cleveref}

\theoremstyle{plain}
\newtheorem{theorem}{Theorem}[section]

\theoremstyle{definition}
\newtheorem{definition}[theorem]{Definition}

\newcommand{\one}{\mbox{$1 \hspace{-1.0mm}  {\bf l}$}}

\newcommand{\bra}[1]{\left\langle{#1}\right\vert}
\newcommand{\ket}[1]{\left\vert{#1}\right\rangle}

\newcommand{\myeps}{\varepsilon}

\newcommand{\eps}[1]{\varepsilon_{\textrm{#1}}}
\newcommand{\leak}{\mathrm{leak}_{\mathrm{EC}}}

\date{\today}

\begin{document}

\title{Finite-key analysis of the six-state protocol with photon-number-resolution detectors}
\author{Silvestre Abruzzo}
\email{abruzzo@thphy.uni-duesseldorf.de}
\author{Markus Mertz}
\author{Hermann Kampermann}
\author{Dagmar Bru{\ss}}
\affiliation{Institute for Theoretical Physics III, Heinrich-Heine-Universit\"at D\"usseldorf, Universitätsstr. 1, 40225 D\"usseldorf, Germany}

\begin{abstract}
 The six-state protocol is a discrete-variable protocol for quantum key distribution, that permits to tolerate a noisier channel than the BB84 protocol. In this work we provide a  lower bound on the maximum achievable key rate of a practical implementation of the entanglement-based version of the six-state protocol. Regarding the experimental set-up we consider that the source is untrusted and the photon-number statistics is measured using photon-number-resolving detectors. We provide the formula for the key rate for a finite initial number of resources. As an illustration of the considered formalism, we calculate the key rate for the setting where the source produces entangled photon pairs via parametric down-conversion and the losses in the channel depend on the distance. As a result we find that the finite-key corrections for the considered scenario are not negligible and they should be considered in any practical analysis.
\end{abstract}

\maketitle

\section{Introduction}
Quantum Key Distribution (QKD) was proposed for the first time in  1984 by Bennett and Brassard\cite{bennett1984quantum}(BB84 protocol) and it is a method for permitting two parties, usually called Alice and Bob, to share a secret bit-string that might be used as a key for cryptographic applications. The most prominent application is encryption with the one-time pad\cite{vernam1926cipher}, where Alice sums bitwise the message and the key for obtaining the cypher-text. The cypher-text is then sent to Bob, who recovers the original text by  using the knowledge of the key. Note that the encrypted text is sent publicly on the channel and therefore it is readable by any eavesdropper who is tapping the channel. The security of this scheme relies on the fact, that from the eavesdropper's point of view the distribution of all possible cypher-texts is uniform\cite{Shannon:1949fk}. This last requirement implies that the key is chosen at random using a uniform distribution on the set of all possible keys. This is the point where QKD  enters the game. In fact, using the laws of quantum mechanics, it is possible to create a bit-string with the guarantee that it is (almost) random from an eavesdropper's point of view\cite{Scarani:2009}.  In this paper we consider the entanglement-based version of the six-state protocol\cite{ekert1991quantum, bennett1992quantum, Brus:1998fk, 6stategisin}. It has been realized that, due to the use of a tomographic measurement, the six-state protocol is more robust against channel imperfections than the BB84 protocol. The six-state protocol was implemented experimentally, e.g. by  Kwiat's group\cite{Kwiat:01}. However, in the meantime the security analysis of this protocol has become more and more complete. In 2001, H.K. Lo\cite{lo2001proof} proved security of the protocol against the most general type of attacks and some years later R. Renner et al.\cite{Renner:2005pi, Kraus:2005kx, Christandl:ye} proved the security of the six-state protocol using information-theoretical arguments. Finite-key effects were considered for the first time by  V. Scarani and R. Renner\cite{Scarani:2008ve, Scarani:2008ys} and by T. Meyer et al.\cite{tim06}. It turns out that there is an initial regime where BB84 is advantageous over the six-state protocol and then there exists a second regime where the six-state protocol leads to higher secret key rates. The reason is the sifting procedure. More precisely, in the standard six-state protocol all measurement bases are chosen with the same probability and as a consequence, $\frac{2}{3}$ of the measurement outcomes are discarded due to this sifting. In the standard BB84 protocol the fraction of discarded outcomes is $\frac{1}{2}$. However, in the year 2005, it was proven by H.K. Lo and M. Ardehali\cite{lo2005efficient} that it is possible to choose one basis with high probability and the other two (one for the BB84) with a negligible probability without jeopardizing the security of the protocol. In the asymptotic case, using this biased scheme, the sifting ratio approaches one and therefore the six-state protocol permits to give a higher secret key rate. However, when finite-key corrections are considered, for small block sizes it is not possible to choose with an arbitrary large bias the measurement basis and therefore the sifting advantage of the BB84 protocol leads to higher key rates.  Note that recent papers considering the finite-key analysis studying the six-state protocol\cite{Scarani:2008ve, Scarani:2008ys, bratzik2010min, abruzzo2011} do not consider a realistic implementation with imperfections in the source, the channel and the detectors. The security proof becomes more involved due to the fact that a realistic source does have multi-photon pulses, which need special care. A common receipt is given by the squashing model\cite{squashing:2008}, which permits to analyze the security of multi-photon sources using single photons and a special post-processing of the outcomes. However, it was proven that an active measurement set-up for the six-state protocol does not permit to use the squashing model\cite{squashing:2008}. A squashing model for the passive measurement set-up exists\cite{squashing:2008}, but up to now only in the case of perfect detectors. However, another technique permitting to overcome the need of squashing model has been developed by T. Moroder et al.\cite{moroder2009}. The main observation is, that if we had perfect photon-number-resolution detectors (PNRD), then we would be able to avoid the problem of multi-photon pulses by post-selecting only single-photon pulses. In their paper the authors developed an experimentally feasible technique permitting to acquire the statistics of a PNRD. In this paper we want to extend their analysis considering finite-key corrections. In order to state clearly our result,  we will consider an ideal set-up, where we perform a Quantum Non-Demolition (QND) measurement permitting to detect error-free the number of photons present in an incoming pulse. We will use a standard measurement set-up, which has detectors with finite efficiency.  Note that, although the set-up we consider may be idealized, it permits to provide a lower bound for the performance of the six-state protocol in presence of a realistic scenario. Finally, we will consider a specific example, i.e. we will calculate the secret key rate in the finite case for a spontaneous parametric down-conversion of type-II (SPDC) source. 

The paper is organized as follows. In section \ref{sec:6state} we describe the  set-up followed by a presentation of the QKD protocol. In section \ref{sec:security} we present the security analysis and the formula for the secret key rate. In section \ref{sec:simulations} we calculate the optimal secret key rate for a SPDC source. Finally, in section \ref{sec:concl} we conclude this analysis. 

\section{The entangled version of the six-state protocol} \label{sec:6state}
In the first part of this section we present the set-up used by Alice and Bob. The second part considers an outline of the QKD protocol. 

\begin{figure}[h]
   \begin{center}

\includegraphics[width=\textwidth]{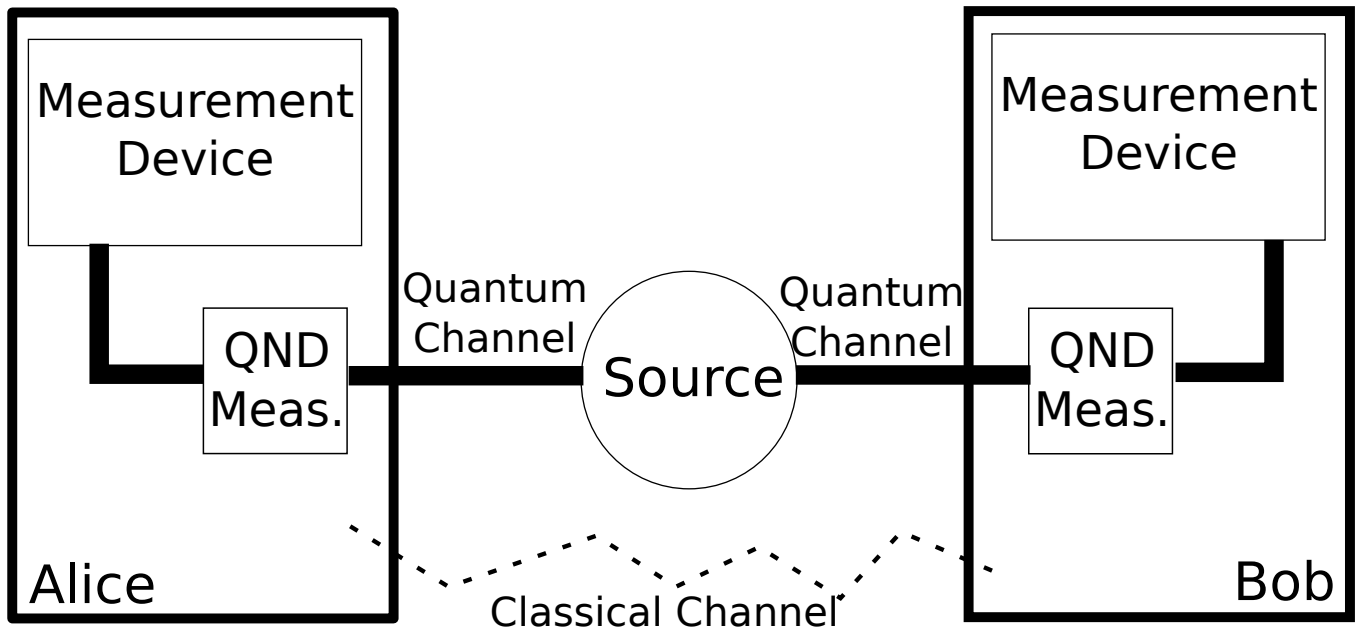}
\caption{\label{fig:set-up} Set-up for QKD. The quantum channel is completely controlled by the eavesdropper. The classical channel is authenticated but otherwise tapped by the eavesdropper. The laboratories are by definition secure.}
\end{center}
\end{figure}

\subsection{Set-up (see \cref{fig:set-up})}

\begin{itemize}

 \item {\bf Source.} An arbitrary source is placed in the middle of Alice and Bob. The source sends an $n$-photon pulse to Alice and an $m$-photon pulse to Bob. 
 \item {\bf Quantum Channel.} We consider that the channel is lossy but otherwise error-free. We suppose that the signals are encoded in the source, such that they do not experience any decoherence in the channel. 
  \item {\bf Classical Channel.} The classical channel is authenticated.
\item {\bf Alice's (Bob's) laboratory.}  We assume that the laboratories are trusted.  Alice (Bob) performs a QND measurement for measuring the number of photons contained in the incoming pulse. The POVM of the  QND measurement is composed of two elements $\{\ket{1}\bra{1}, \one-\ket{1}\bra{1}\}$, where $\{\ket{n}\}$ is the Fock-basis. After the QND measurement, the pulse passes a standard QKD-measurement set-up, where one measurement basis is chosen at random out of the $X$-, $Y$- and $Z$-direction. Note that regarding the detectors, we assume that they have finite efficiency $\eta_{D}$ and negligible background noise. Moreover, we consider a misalignment\cite{moroder2009,Scarani:2009} in the detectors. Each time that a single photon arrives at the detection device, it is measured correctly with probability $1-\eta_{M}$.
\end{itemize}

\subsection{QKD protocol}\label{subsec:QKDprot}
\begin{enumerate}
\item{ \bf Entanglement generation and distribution.} A source generates entangled pairs which are distributed through the quantum channel to Alice and Bob.

 \item {\bf Measurement.} Alice and Bob choose at random and independently the measurement basis and to perform the measurement on the incoming pulse. We consider a biased choice of the bases, i.e., the basis $Z$ is chosen with probability $p_{Z}\geq \frac{1}{3}$ and the other two bases are chosen with the same probability $p_{X}=p_{Y}$. The result of the measurements is recorded in a vector of the form $(t^{A},b_0^{A}, b_1^{A}, p^{A}, basis^{A})$, where $b_i^{A}=0$ indicates that the detector for the classical value $i$ on Alice side did not experience a click, otherwise $b_i^{A}=1$. The entry $p^{A}$ contains the result of the QND measurement, in particular $p^{A}=1$ when a 1-photon pulse is measured and $p^{A}=0$ otherwise. The last entry contains a label for the measurement basis. The first entry $t^{A}$  is a tag permitting to distinguish the measurements, e.g., the time of occurrence of the measurement.

\item {\bf Vacuum sifting.} During this sifting we remove the non-measurement results. This step is performed locally and without communication between Alice and Bob. Let i=A,B.  When $p^{i}=1$, it is still possible that $b_0^{i} + b_1^{i}=0$, i.e., none of the detectors has clicked. This can happen due to the finite efficiency of the detectors. We can eliminate these events safely, incorporating the efficiency of the detectors in the efficiency of the channel. During this step Alice (Bob) calculate the value of $b_0^{i}+ b_1^{i}$ and set $p^{i}=0$  every time that  $b_0^{i} + b_1^{i}=0$.

 \item {\bf Pulse sifting.} We use the output of the QND measurement for conditioning the type of bits used for the key. Alice and Bob communicate via the classical channel the value of $p^{A}$ and $p^{B}$ for each measurement and discard all measurements with $p^{A}\times p^{B}\neq1$\cite{moroder2009}. Note that this post-processing is possible only due to the fact, that the QND measurement is perfect and that we are considering entanglement-based QKD. If one of the two assumptions above is dropped, then security loopholes will arise\cite{Scarani:2009}.
 
\item {\bf Bases sifting.} Alice and Bob exchange information regarding the measurement bases and discard the outcomes coming from different bases.

\item {\bf Parameter estimation.} Alice and Bob take a random sample from each basis and use  this sample for estimating the Quantum Bit Error Rate (QBER) for each basis. We denote with $e_{X, m_{X}}$ the fraction of  erroneous  bits in the sample of length $m_{X}$. We choose\cite{bratzik2010min} $m_{X}=m_{Y}=m_{Z}:=Np_{X}^2$ , where $N$ is the number of bits after the pulse sifting. The QBERs along the Y and Z bases are defined analogously. Note that the worst-case QBER can be estimated with the fluctuations due to the finiteness of the sample.

\item {\bf Error correction.} During this step Alice and Bob apply a one-way error correction protocol  and correct their strings. As result they will exchange $\textrm{leak}_{\textrm{EC}}$ bits on the channel.

\item {\bf Error verification.} In realistic implementations it is possible that at the end of the error correction protocol, Alice and Bob do not have perfectly correlated bits. In order to acquire confidence regarding the remaining errors, they apply a two-universal hash function on their strings and they communicate the result of the function on the channel.
 This step costs $\log_2(\frac{2}{\eps{EC}})$ bits. If the resulting hash tag is the same, then the two strings are the same with probability $1-\eps{EC}$ . If the hashing produces a different outcome, Alice and Bob may perform more error correction followed by another error verification.

\item {\bf Privacy amplification.} Alice and Bob apply a two-universal hash function in order to shrink their string. The resulting string is called the key. \\
In the next section we will discuss a bound on the achievable key length $\ell$ as a function of a security parameter $\eps{}$.
\end{enumerate}

\section{Finite secret key rate}\label{sec:security}
The secret key rate is the relevant figure of merit for describing the performance of a QKD protocol. First of all, we are going to state the definition of security.

\begin{definition}\label{def:securitydef}\cite{renner05, muller2009composability} Let $\rho_{KE}$ be the classical-quantum-state describing the classical key $K$ of length $\ell$, distilled at the end of a QKD protocol, correlated with the quantum states of the eavesdropper $\rho_{E}$. The state $\rho_{KE}$  is said to be $\myeps$-secure if
\begin{equation}
\label{def:security}
\underset{\rho_{E'}}{\mathrm{min}}\frac{1}{2}\|\rho_{KE}-\frac{1}{2^\ell}\one\otimes\rho_{E'}\|_{1}\leq\myeps,
\end{equation}
where $\rho_{E'}$ is the quantum state of an eavesdropper not correlated with the key.
\end{definition}

The definition states that from the eavesdropper's (Eve) point of view the classical key $K$ is indistinguishable from a random and uniform key with probability $1-\epsilon$.  Note that the used definition of security is composable, i.e. if we have two protocols characterized by two different probabilities of failure, then, after a concatenation of these protocols, the probability of failure of the global protocol will be bounded by the sum of the single probabilities of failure.

In the following we derive a formula for the $\myeps$-secure key length $\ell$. We consider that Eve has complete control over the quantum channel and the source. Moreover,  we consider the \emph{uncalibrated scenario}\cite{Scarani:2009}, where the finite efficiency of the detectors are also attributed to Eve.
\newcommand{\nsource}{N_{\mathrm{source}}}
Let $p_{11}$ be the probability that Alice and Bob receive a single photon. Then, starting with $\nsource$ initial pulses, the steps $1-4$ of the QKD protocol (see \cref{subsec:QKDprot}) decrease the number of signals to $\nsource p_{11}$. Afterwards, the bases-sifting and the PE lead to $\nsource p_{11}\left(p_{Z}^2-p_{X}^2\right)$ resulting bits. 
For PE $3p_{X}^{2}$ signals are used to estimate the QBER. The fluctuations due to finite statistics have been analyzed in \cite{Scarani:2009wr,Scarani:2008ve,Scarani:2008ys,bratzik2010min} . Note that differently to \cite{bratzik2010min} we do not consider one symmetrized QBER. Instead we treat the QBER for each direction separately.

Let $e_{i, m_{i}}$ be the measured QBER in direction $i=X,Y, Z$, then with probability $1-\myeps_{PE}$ the real QBER $e_{i}$ is such that\cite{Scarani:2009wr,Scarani:2008ve,Scarani:2008ys,bratzik2010min}
\begin{equation}
\label{eq:pa:boundQBER}
 e_i \leq  e_{i, m_{i}} +  2\zeta\left(\myeps_{PE},m_{i}\right)
\end{equation}
with
\begin{equation}
\zeta(\myeps_{PE},m):=\sqrt{\frac{\ln{\left(\frac{1}{\myeps_{PE}}\right)}+2\ln{(m+1)}}{8m}}.
\end{equation}

For the error correction protocol the total number of bits exchanged during the procedure is an upper bound on the information leaked to the eavesdropper about the final key.  For the simulations, we will use   \cite{Scarani:2009wr, Scarani:2008ve}
\newcommand{\fec}{f_{\mathrm{EC}}}
 \begin{equation}
\label{eq:leakec}
 \leak := f_{\mathrm{EC}}nh(e),
\end{equation} where $f_{\mathrm{EC}}\geq1$ depends on the used EC protocol, $h(e)$ is the binary Shannon entropy, i.e.,  $h(e)=-e\log{e}-(1-e)\log{(1-e)}$ and $e$ is the QBER.  This definition comes from the fact that $nh(e)$ represents the asymptotic number of bits used by a perfect error correction protocol. The coefficient $f_{\mathrm{EC}}$ represents a deviation of the real protocol from the asymptotic one. 

Regarding privacy amplification many bounds on the achievable secret key length are placed at the disposal in the literature\cite{Scarani:2008ve,Scarani:2008ys,bratzik2010min,abruzzo2011}. Note that the bounds given in \cite{bratzik2010min,abruzzo2011} are tighter than the bound given in \cite{Scarani:2008ve,Scarani:2008ys} . However, they require that the channel is symmetric. Although it is possible to transform  any channel in a symmetric one, we consider the bound provided in \cite{Scarani:2008ve,Scarani:2008ys} to take the analysis simple and more general.

The following result summarizes the preceding considerations and provides a formula for the achievable secret key length. It is important to emphasize, that the following theorem holds only due to our special set-up with the QND measurement and the particular post-processing, which selects only the pulses containing one photon.

\begin{theorem}[\cite{Scarani:2008ve,Scarani:2008ys}]
 Let $\nsource$ being the number of measurements performed by Alice and Bob. Let $p_{11}$ be the fraction of attempts resulting in a single-photon pulse entering Alice's and Bob's laboratories. The number of bits allocated for extracting a key is $n:=\nsource p_{11}(p_{Z}^2-p_{X}^2)$. If Alice and Bob distill a key of length
\begin{equation}\label{eq:keyrate}
 \ell\leq \underset{\overline{\eps{}},\eps{PE},\eps{PA},p_{X}, p_{11}}{max}\left[n(S_\zeta(X|E)-\fec h(e_{Z}))-2\log_{2}{\frac{1}{\eps{PA}}}-\log_{2}{\frac{2}{\eps{EC}}}\right],
\end{equation}
then it is $\eps{}$-secure, with $0\leq\overline{\eps{}}+\eps{EC}+\eps{PA}+\eps{PE}\leq\eps{}$. The quantity $S_\zeta(X|E)$ is given by \cite{Scarani:2008ve,Scarani:2008ys, Scarani:2009}
\begin{equation}
 S_\zeta(X|E):=1-e_{Z}h\left(\frac{1+(e_{X}-e_{Y})/e_{Z}}{2}\right)-(1-e_{Z})h\left(\frac{1-(e_X+e_Y+e_Z)/2}{1-e_Z}\right)-5\sqrt{\log_{2}\left(\frac{2}{\overline{\eps{}}}\right)\frac{1}{n}}.
\end{equation}
\end{theorem}

The entropy $S_\zeta(X|E)$ is calculated with the QBER inferred during the parameter estimation protocol (see \cref{eq:pa:boundQBER}). We would like to point out that the theorem above is a standard theorem, the unique difference is that we are not using all signals for extracting the key but only the signals coming as single-photon pulse. 

The asymptotic formula for the secret key rate can be recovered as a special case of the theorem above for $n\rightarrow\infty$ and $\eps{}\rightarrow0$.

\section{Case study: SPDC source}\label{sec:simulations}
In this section we will calculate the achievable secret key length for a pumped type-II down-conversion source\cite{spdckok}. The produced state by this source can be written as
\begin{equation}
\label{eq:state:pdc}
 \ket{\phi}_{AB}:=\sum_{n=0}^{\infty}\sqrt{p_{n}}\ket{\phi_n}_{AB},
\end{equation}
where 
\begin{equation}
 p_{n}:=\frac{(n+1)\lambda^n}{(1+\lambda)^{n+2}},
\end{equation}
and 
\begin{equation}
\label{eq:state:pdcn}
 \ket{\phi_{n}}_{AB}:=\sum_{m=0}^{n}\frac{(-1)^m}{\sqrt{n+1}}\ket{n-m, m}_{A}\ket{m, n-m}_{B}.
\end{equation}
The state above is written along one fixed direction, e.g. the $Z$-direction. The meaning of the notation $\ket{l_{H}, l_{V}}_{A}$ is that on Alice side, a pulse with $l_{H}+l_{V}$ photons is coming and $l_{H}(l_{V})$ have horizontal (vertical) polarization.

The quantity $2\lambda$ represents the mean photon pair number per pulse. 

In the following we calculate the quantities that enter the formula of the secret key rate (\cref{eq:keyrate}). First of all, we express the probability that Alice and Bob receive only one photon. Then we calculate the QBER produced by the incoming pulse and finally, we find the optimal mean photon pair number per pulse, i.e. the one which maximize the secret key rate.

\subsection{Calculation of $p_{11}$}

We denote with $\eta_{A}$ the total transmittivity of Alice's set-up. It is given by $\eta_{A}:=\eta_{D}\eta_C(L/2)$, where $\eta_{D}$ is the efficiency of Alice's detectors and $L$ is the distance between Alice and Bob. We consider a lossy, but otherwise perfect channel with attenuation coefficient $\alpha=0.17$ dB/km, such that the transmission probability of a photon is given by $\eta_{C}(L):=10^{-\frac{\alpha L}{10}}$.

Analogously we define the total efficiency on Bob's set-up, denoted by $\eta_{B}$.
When an $n$-photon pulse is produced, during its travel on the channel and during the detection, some photons could be absorbed. The following formula gives the probability that an $n$-photon pulse becomes a $1$-photon pulse,
\newcommand{\etaA}{\eta_{A}}
\newcommand{\etaB}{\eta_{B}}
\begin{equation}
 W_{n}:=p_{n}n^2(1-\etaA)^{n-1}(1-\etaB)^{n-1}\etaA\etaB.
\end{equation}
The factor $n^2$ is a combinatorial factor coming from our ignorance which photon was absorbed. The total probability that both, Alice and Bob, receive one photon is given by 
\begin{equation}
\label{eq:p11pdc}
p_{11}:=\sum_{n=1}^{\infty}W_{n}.
\end{equation}

\subsection{Calculation of the QBER}
In the six-state protocol measurements are performed along three orthogonal directions in the Bloch sphere, and, as explained above, three QBERs are involved. The Hamiltonian of the parametric down-conversion process is invariant under rotations from the X- to Y-, Y- to Z- and Z- to X-basis. Therefore, the state in \cref{eq:state:pdc} remains invariant in form under these transformations and hence the QBER is the same in all directions, i.e., there is only one QBER to consider, e.g., for the Z-direction. There are two contributions to the QBER. The first one comes from the misalignment and the second one is due to the fact, that the entering state is not maximally entangled. Let $e_{n}$ be the QBER generated by $\ket{\phi_n}$ when misalignment is not considered. Then the total QBER is given by
\begin{equation}
\label{eq:epdc}
 e_{PDC}:=\frac{\sum_{n=1}^{\infty}(e_{M}(1-e_{n})+(1-e_{M})e_{n})W_{n}}{p_{11}},
\end{equation}
where $e_{M}:=2\eta_{M}(1-\eta_{M})$ and $\eta_{M}$ is the misalignment-error probability. The first term of $e_{PDC}$ accounts for the fact, that even if the incoming state did not produce a QBER, due to the misalignment there would be an error. The second contribution comes from the error generated by the incoming photons. Note that terms of the form $e_Me_n$ are not considered, because the simultaneous appearance of these two errors will produce correlated outcomes. The quantity $e_{n}$ can be calculated with the help of \cref{eq:state:pdcn}. This state is the superposition of $n+1$ states. The first and the last term in the summation, with $m=0$ or $m=n$ will produce a correlated outcome. On the contrary the remaining $n-1$ elements in the summation will produce an error with probability $\frac{1}{2}$. Therefore we get
\begin{equation}
 e_n=\frac{(n+1)-2}{2(n+1)}=\frac{1}{2}\left(1-\frac{2}{n+1}\right).
\end{equation}
From the formula above it is possible to verify that $e_1=0$, which is consistent with the fact that $\ket{\phi_{1}}$ is a maximally entangled state. 

The common free parameter in the QBER $e_{PDC}$ and in $p_{11}$ is the mean number of photons per pulse $2\lambda$. 

Therefore, in the following we will calculate the optimal $\lambda$ permitting to maximize the secret key rate. 

\begin{figure}[h]
\begin{center}

\includegraphics[width=14cm]{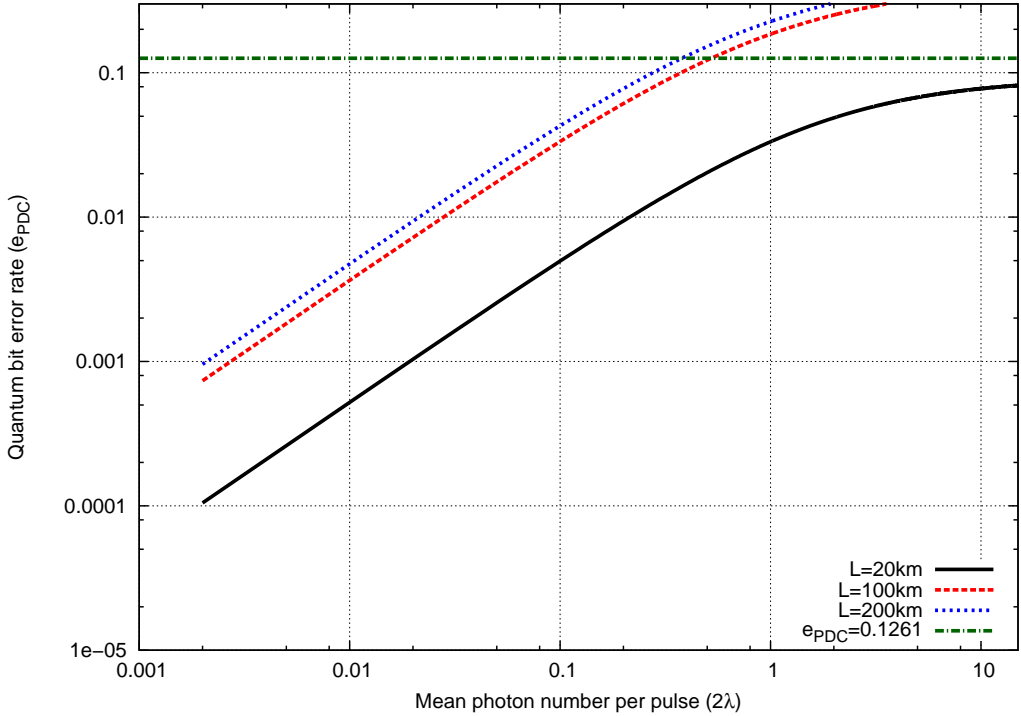}
\caption{\label{fig:epdc}(Color online) Value of $e_{PDC}$ (\cref{eq:epdc}) as a function of the probability that both, Alice and Bob, receive one photon as a function of the mean number of photons produced by the source for various distances. The horizontal line represents the maximal QBER tolerated by the six-state protocol. The absorption of the channel is $\alpha=0.17$ dB/km and Alice and Bob use perfect detectors $\eta_{D}=1, \eta_{M}=0$.}
\end{center}
\end{figure}

As shown in \cref{fig:epdc}, in order to have a low QBER $e_{PDC}$ it is necessary to have $\lambda$ small. For short distances, e.g. $L=20$km it is possible to choose $\lambda< 20$ and at the same time be able to extract a key. The reason is that the multi-photon pulses arrives to Alice and Bob without an appreciable degradation and therefore, we are able to filter those contribution to the QBER during the pulse sifting. However, the situation changes when  the distance between Alice and Bob increases. We see that the mean number of photons per pulse has to be much smaller than $1$ in order to decrease the multi-photon contribution to the QBER. From \cref{fig:epdc} we see that for $L\geq100$km we have to choose $\lambda<1$ in order to have a QBER smaller than the maximal QBER tolerated by the six-state protocol. 

\subsection{Asymptotic secret key rate}

The secret key rate in the asymptotic case characterizes the maximal achievable secret key rate in case of perfect error correction, no uncertainty in the estimation of the QBER and perfect security ($\epsilon=0$). The formula is given by
\begin{equation}\label{eq:asymrate}
 r_{\infty}:=\lim_{\substack{
            n\rightarrow\infty\\
            \eps{}\rightarrow0}}\frac{l}{\nsource}=\underset{\lambda}{\textrm{max}}\left[p_{11}\left(\left(1-e_{PDC}\right)\left(1-h\left(\frac{1-3e_{PDC}/2}{1-e_{PDC}}\right)\right)-h\left(e_{PDC}\right)\right)\right].
\end{equation}

\begin{figure}[h]
   \begin{center}
\includegraphics{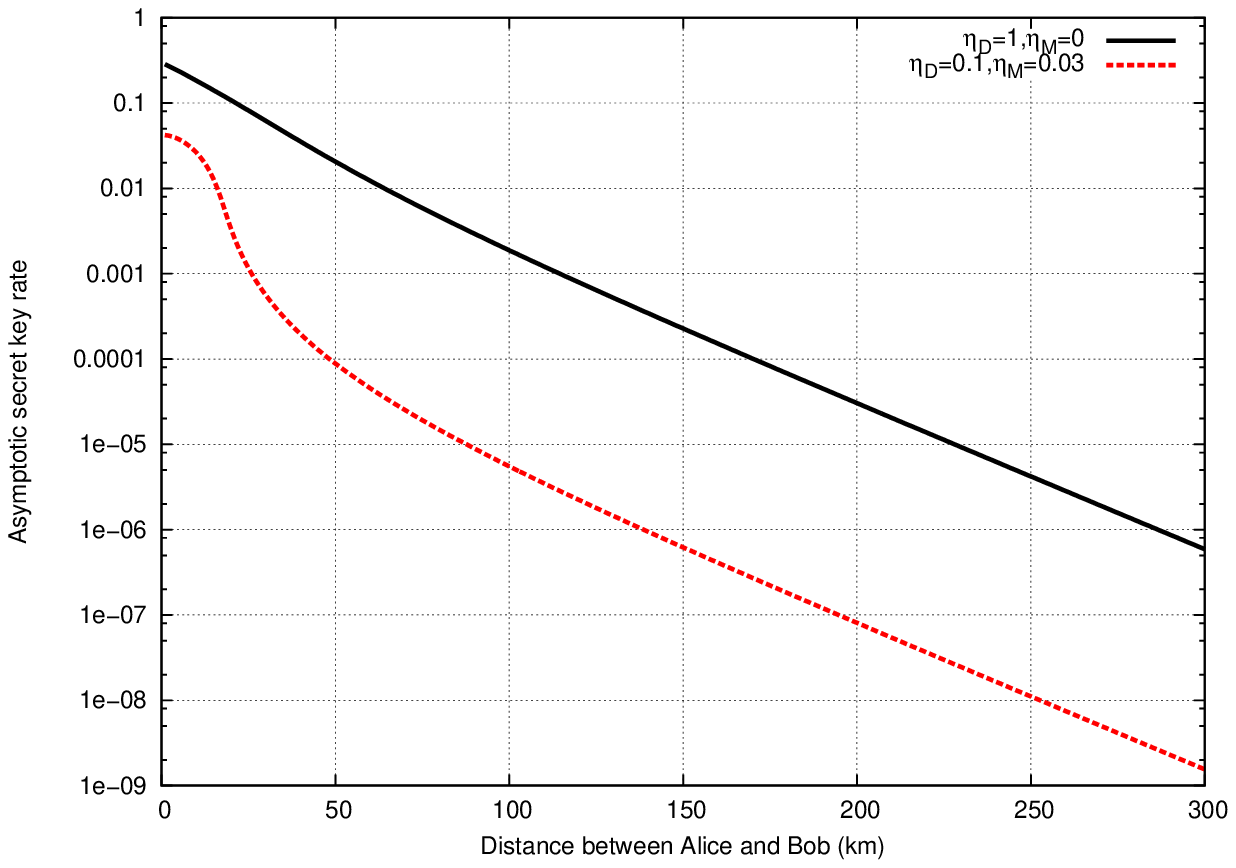}
\caption{\label{fig:krasym}(Color online) Asymptotic secret key rate (\cref{eq:asymrate}). The absorption of the channel is $\alpha=0.17$ dB/km.}
\end{center}
\end{figure}

\begin{figure}[h]
\begin{center}
\includegraphics{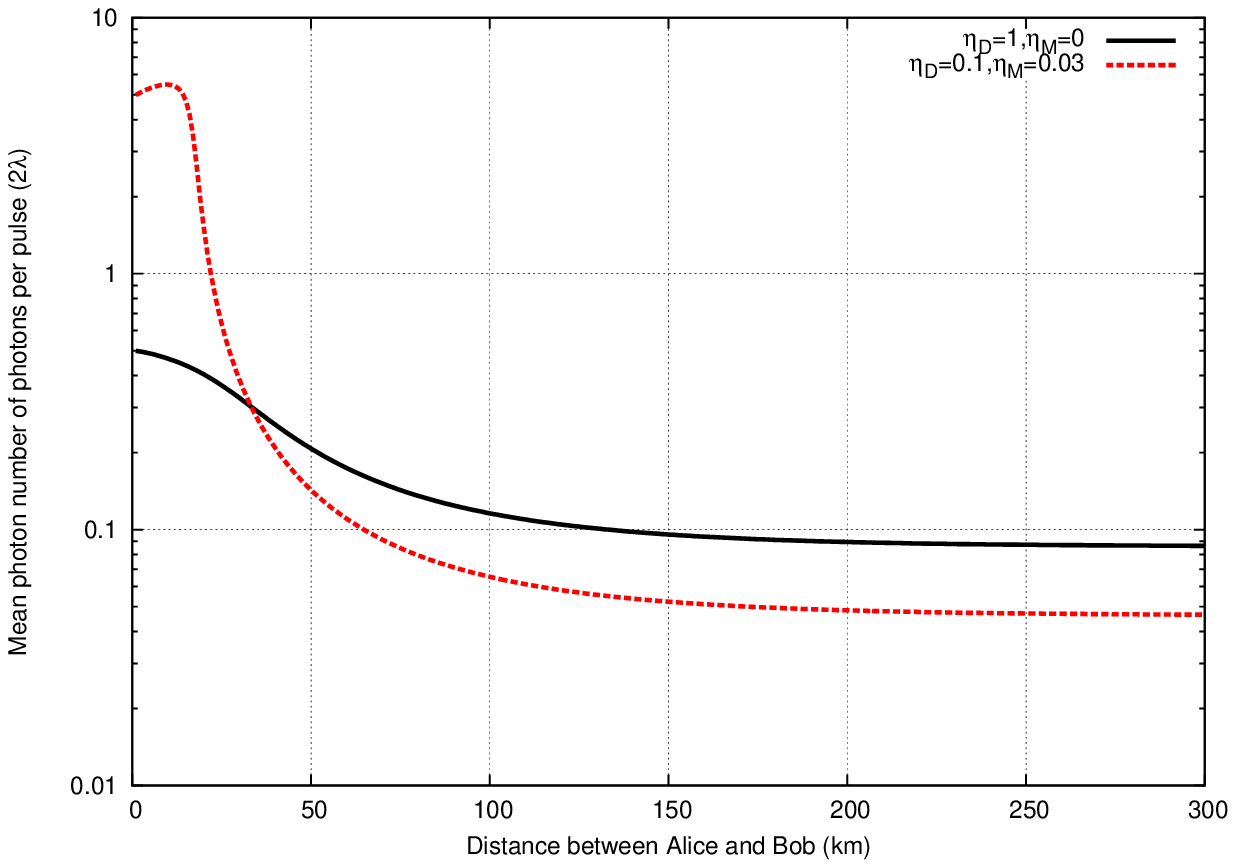}
\caption{\label{fig:krasymptlambda}(Color online) Optimal mean of photon-number per pulse. The absorption of the channel is $\alpha=0.17$ dB/km.}
\end{center}
\end{figure}

In \cref{fig:krasym} the secret key rate is shown as a function of the distance for two different experimental set-ups. A comparison between an idealized scenario ($\eta_{D}=1, \eta_{M}=0$) and a more realistic one ($\eta_{D}=0.1, \eta_{M}=0.03$) shows that the secret key rate decreases of at least $2$ orders of magnitude. Regarding the optimal mean of photon-number per pulse, as shown in \cref{fig:krasymptlambda}, the difference is of the order of $1$. The optimized function is non-linear and the used optimization algorithm may only permit to find a local optimum. 

Finally, we would like to point out, that a similar analysis of the asymptotic case was performed by Moroder et al.\cite{moroder2009} with a source placed in an asymmetric position, i.e., closer to Bob than to Alice.

\subsection{Finite-key analysis}
In a practical execution of a QKD protocol, the initial number of resources is always finite, therefore we need to take into account corrections to the asymptotic secret key rate. The formula for the secret key rate is
\begin{align}
 r:=\frac{\ell}{\nsource}=\underset{\overline{\eps{}},\eps{PE},\eps{PA},p_{X}, \lambda}{max}&\left[p_{11}(p_{Z}^2-p_{X}^2)\left(\left(1-e_{PDC}\right)\left(1-h\left(\frac{1-3e_{PDC}/2}{1-e_{PDC}}\right)\right)-\fec h(e_{Z})\right)\right.\\
&\left.\quad\quad\quad-2\log_{2}{\frac{1}{\eps{PA}}}-\log_{2}{\frac{2}{\eps{EC}}}-5\sqrt{\log_{2}\left(\frac{2}{\overline{\eps{}}}\right)}\right].
\end{align}
The calculations are done in such a way, that we optimize over all free parameters: the mean number of photons per pulse ($\lambda$), the probability to measure along the $Z$ basis ($p_{Z}$), the failure probability for the parameter estimation ($\eps{PE}$), for privacy amplification ($\eps{PA}$) and the smoothing parameter ($\overline{\eps{}}$).

\begin{figure}[h]
\begin{center}
\includegraphics{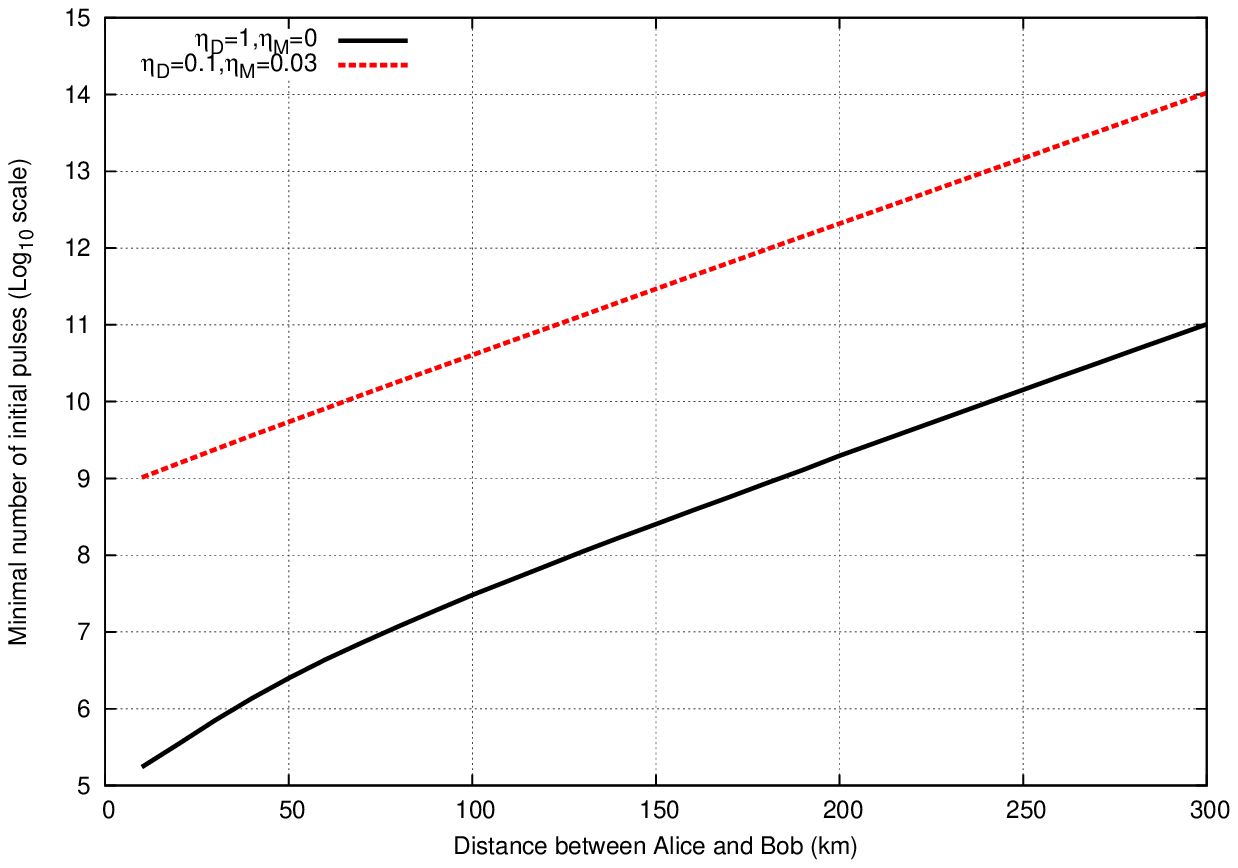}
\caption{\label{fig:MinSPDC}(Color online) Minimal number of initial pulses permitting to extract a key of 1 bit as a function of the length $L$ for a perfect set-up ($\eta_D=1$, $\eta_{M}=0$). The absorption of the channel is $\alpha=0.17$ dB/km. Security parameter $\eps{}=10^{-9}$, $\eps{EC}=10^{-10}$, $f_{EC}=1.2$.}
\end{center}
\end{figure}

For extracting a key it is necessary to have a block bigger than a specific length. As shown in \cref{fig:MinSPDC}, even for short distances the source has to emit at least $10^5$ pulses with in mean $\lambda\approx0.1$ photons per pulse for extracting a key of 1 bit. However, if we consider detector inefficiencies and misalignment errors, the requirements will become much more stringent. In particular, we need at least $10^9$ pulses for extracting a key.

\begin{figure}[h]
   \begin{center}

\includegraphics{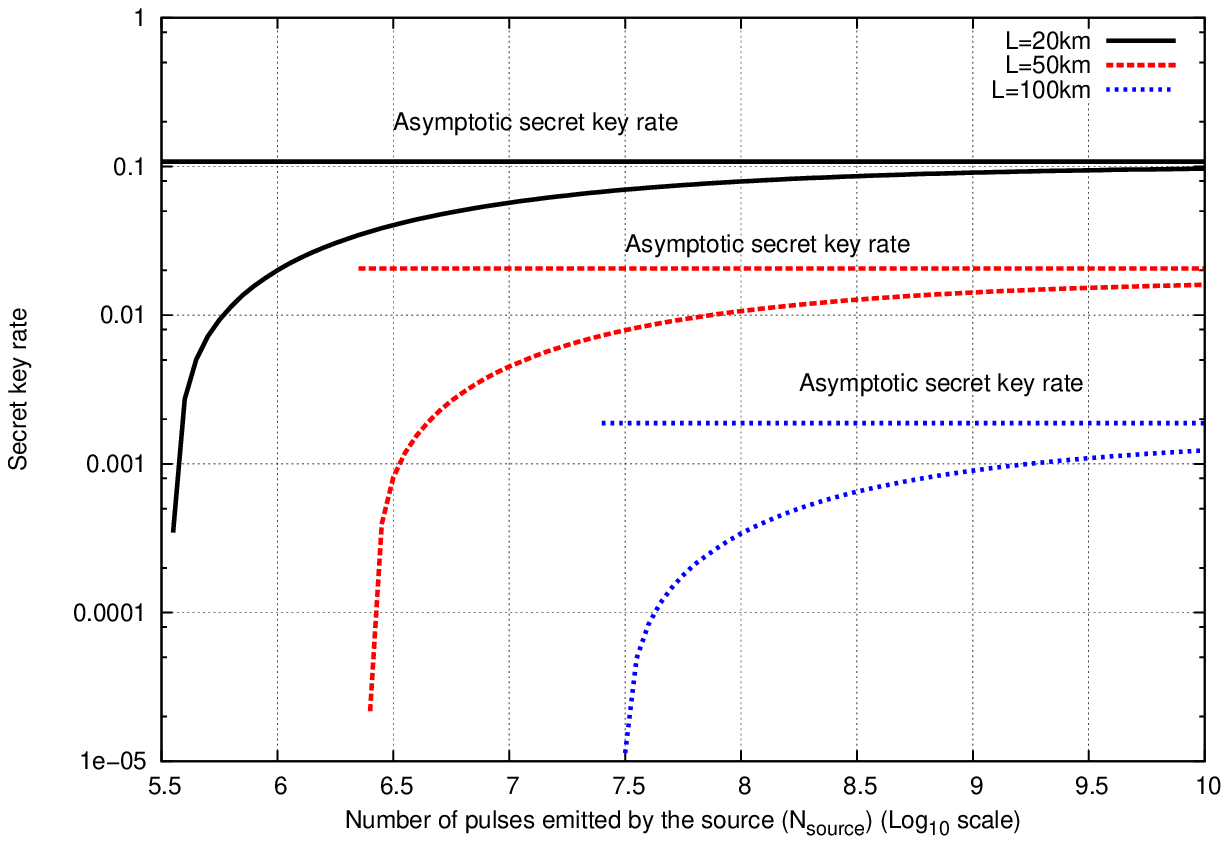}
\caption{\label{fig:krfinperf}(Color online) Secret key rate as a function of the number of pulses emitted by the source ($\nsource$) for a perfect set-up ($\eta_D=1$, $\eta_{M}=0$). The absorption of the channel is $\alpha=0.17$ dB/km. Security parameter $\eps{}=10^{-9}$, $\eps{EC}=10^{-10}$,$f_{EC}=1.2$.}
   \end{center}
\end{figure}

The second quantity we want to analyze is the secret key rate (\cref{eq:keyrate}). As shown in \cref{fig:krfinperf}, for a perfect set-up ($\eta_D=1$, $\eta_{M}=0$) the finite secret key rate differs significantly from the asymptotic secret key rate. In particular, for all distances considered in \cref{fig:krfinperf}, the secret key rate differs of at least 10\% ($\nsource=10^{10}$, $L=20$ km) from the asymptotic key rate. However, for more realistic initial number of pulses, the difference is bigger, e.g for $L=100$ km and $\nsource=10^8$, the difference between the asymptotic secret key rate and the one with finite-key corrections is of one order of magnitude. In case of imperfections we will have similar plots but with a worse secret key rate. However, the qualitative behavior of the plot remains similar to \cref{fig:krfinperf}.

\section{Conclusion}\label{sec:concl}
In this paper we did a step towards the analysis of a realistic implementation of  the entanglement-based version of the six-state protocol. We considered that the standard QKD measurement is preceded by a QND measurement permitting to know the number of photons entering in the source. This special set-up with a post-processing which considers only signals coming from a single-photon source permits to evaluate secret key rates for the six-state protocol. We studied the case of an arbitrary large number of initial pulses as well as of a finite key. As result we found that in realistic implementations with finite-efficiency detectors and misalignment, the minimal number of pulses for being able to extract a key is around $10^9$ pulses at the distance of a few kilometers. Note that this is a very stringent requirement. In fact, considering an ordinary source, which emits pulses at the rate of $10$ MHz, at the distance of $20$ km between Alice and Bob, the time needed for extracting a key of 1 bit will  be of the order of 100 seconds. Using the asymptotic key formula, in the same time, it could be possible to obtain a key of length $10^6$ bits, which would be unfortunately completely insecure. Therefore, we emphasize once again that finite-key corrections are necessary for a realistic and correct security analysis.

Regarding future work, we underline that more realistic experimental imperfections should be taken into account in order to characterize the performance of the six-state protocol. In a future work, we want to consider the encoding of the quantum bits on the quantum channel and to study the effects of decoherence. This is a problematic issue which limits practical implementations of the six-state protocol and needs a careful analysis.

\section*{Acknowledgments} 
We would like to thank Sylvia Bratzik and Tobias Moroder for valuable and enlightening discussions. We acknowledge partial financial support by Deutsche Forschungsgemeinschaft (DFG) and by BMBF (project QuOReP).

\bibliography{finitekey}
\bibliographystyle{apalike}
\end{document}